\documentclass[usenatbib]{mn2e}
\newcommand{\msun}{$M_\mathrm{\sun}$ }

\usepackage{graphicx}

\title[Disruption of a Dwarf Galaxy Under Strong
  Shocking: \\ The Origin of $\omega$ Centauri]{Disruption of a Dwarf Galaxy Under Strong
  Shocking: \\ The Origin of $\omega$ Centauri}
\author[T. Tsuchiya, V. I. Korchagin and D. I. Dinescu]
  {T. Tsuchiya$^{1,2}$\thanks{e-mail: tsuchiya@ari.uni-heidelberg.de},
  V. I. Korchagin$^{3,5}$ and D. I. Dinescu$^{3,4}$ \\
  $^{1}$Astronomisches Rechen-Institut, M\"onchhofstra\ss e 12-14, 69120
  Heidelberg, Germany \\
  $^{2}$ SGI Japan, Ltd., Yebisu Garden Place Tower, 4-2-3 Ebisu Shibuya-ku, Tokyo 150-6031, Japan\\
  $^{3}$Astronomy Department, Yale University,\\
  $^{4}$Astronomical Institute of The Romanian Academy\\
  $^{5}$Institute of Physics, Rostov-on-Don, Russia
}

\begin{document}
\date{Accepted ----. Received ----.}
\pagerange{\pageref{firstpage}--\pageref{lastpage}} \pubyear{2002}
\maketitle
\label{firstpage}

\begin{abstract}

We perform N-body simulations of the dynamical evolution of a dwarf galaxy falling
into the Milky Way  galaxy in order to understand the formation scenario 
of the peculiar globular cluster $\omega$ Centauri. We use  self-consistent models of
the bulge and the disc of the Milky Way,
as well as of the dwarf galaxy, and explore a range of dwarf models
with different density distributions. Namely, we use King  (1966)
and Hernquist (1990) density profiles to model the density distribution
in the dwarf. The central region of our King model has a density profile approximately
$\propto r^{-2}$, while that of the Hernquist model is $\propto r^{-1}$. 
The difference in the dwarf's density distributions leads
to distinct evolutionary scenarios. The King model dwarf loses its mass exponentially
as a function of apocentric distance, with the mass-loss rate depending
on the initial mass and size of the dwarf. Regardless of the initial
mass and size, the King model dwarf remains more massive than $10^8$ \msun
after a few  Gyr of evolution. The Hernquist model dwarf experiences an 
accelerated mass loss, and the mass of the remnant falls below 
$10^8$ \msun within a few Gyr. By exploring an appropriate set of parameters,
we find a Hernquist model that can attain the mass and 
orbital characteristics of $\omega$ Cen after a few Gyr.    

\end{abstract}


\section{Introduction}
\label{sec:Introduction}

The accretion and subsequent tidal disruption of dwarf galaxies are
important agents in the formation and evolution of the Milky Way
(Searle \& Zinn 1978). Such accretion events would presumably 
lead to formation of tidal streams in the halo of the Milky Way
associated with the disrupted dwarf galaxies.
The discovery of the tidal stream related to the Sagittarius
dwarf galaxy (Ibata et al. 2001) indeed supports the idea
that halos of galaxies can have a number of
streams maintaining their coherence during a few Gyrs.

There is another class of halo objects which might be associated with
past accretion events.
During the accretion of nucleated dwarf elliptical galaxies,
 the compact stellar nuclei may 
survive the disruptive event and continue orbiting until the present time.
Freeman \& Bland-Hawthorn (2002) even speculate that the
globular clusters in the Milky Way could be the stripped relics of ancient
protogalactic stellar systems. To identify such objects
would be of great importance in understanding the nature
of such building blocks of galaxies.
In this paper we focus on one object which
may be the relic of a past accretion event - the
globular cluster $\omega$ Centauri.
$\omega$ Centauri is one of the most peculiar
globular clusters in the Milky Way:
with a mass of $5\times10^6$\msun, it is the most massive globular cluster.
It has an unusual flatness with an ellipticity of about 0.12 associated with
fast internal rotation  \citep{free2001}. 
It orbits in a retrograde direction relative to the disc
with an apocentric radius of about 6 kpc, a pericentric radius
of 1.2 kpc, and a maximum height above the plane of 1 kpc
\citep{dine1999}.

What really distinguishes $\omega$ Cen from other
Galactic globular clusters, is its chemical abundance pattern.
It is the only known globular cluster that has well-defined
signatures of self-enrichment. Thus, according to Smith et al. 2000
and Vanture et al. 2002, the abundance pattern observed within $\omega$ Cen
can be understood as a combination of Type II supernovae and,
more importantly, AGB-star enrichment during a period of about 3 Gyr.
Gnedin et al. (2002) have demonstrated that $\omega$ Cen on its present
orbit and with its present mass can not retain heavy elements dispersed in the AGB phase of stellar
evolution. The passages through the disk would sweep out the intracluster gas
and the cluster, once established on this orbit, would not 
be able to produce stars with enhanced s-process elements.
Thus, the cluster should have evolved chemically
before settling on its present orbit, or, in other words, it was
accreted by the Galaxy.
Another possibility to explain the chemical peculiarities of $\omega$ Cen
is the merger of two or more chemically distinct systems.
Vanture et al. (2002) conclude however, that in this scenario, it would be
difficult to explain an increase in the of ratio of heavy s-process elements
to iron with metallicity.   

A plausible explanation is that   $\omega$ Cen
is the nucleus of a dwarf galaxy captured and disrupted 
by the the Milky Way (e.g., Freeman 1993).
In this scenario, a dwarf sinks to the center of
the Galaxy due to dynamical friction,  simultaneously losing 
mass in the Galactic tidal field. All of the dwarf's mass is eventually
stripped but the nucleus, with a mass of $5\times10^6$ \msun, is
left on the current orbit of $\omega$ Cen.
The current angular momentum of $\omega$ Cen is thus
inherited from the dwarf galaxy, on a retrograde orbit.
The complex chemical composition and extended star formation history
of $\omega$ Cen can be explained
in this picture by the chemical evolution of the nucleus
in a deep potential well of a dwarf, such that the nucleus would be able to
retain stellar ejecta during subsequent activity of supernovae and AGB stars.

Even though a capture scenario is seen as a plausible explanation for the
peculiar properties of $\omega$ Cen, it is unclear
whether it is possible to realize an orbit and mass evolution of the dwarf
that can reproduce the current parameters of $\omega$ Cen.  
Zhao (2002) studied the orbital decay of a dwarf
that may lead to the current position of $\omega$ Cen by
launching a dwarf on an orbit that started 50 kpc away from the Galactic center.
Using a semi-analytical model, he concluded that a progenitor
of $\omega$ Cen can not decay to its present orbit.  
He found that strong tidal shocks
quickly reduce the mass of the dwarf so that dynamical friction becomes too weak
to drag the remnant to the inner regions of the Galaxy. 
Zhao (2002) concluded that the only possibility to explain 
$\omega$ Cen - phenomenon is that its progenitor was born 
only $15$~kpc away from the Galactic center.

Zhao's analysis, however,  was simplified in a number of ways.
He assumed that the mass outside the tidal radius is
instantaneously stripped,  and the contribution from the bulge 
and disc components to the dynamical drag was not taken into account. 
In this study, we perform a series of numerical simulations,
using self-consistent bulge and disc models for the Milky Way, and 
self-consistent dwarf models that have different density profiles. 
As a result, we find a dwarf-capture scenario that successfully
produces an $\omega$ Cen-like object from a normal dwarf
galaxy. This result is also reported in more detail in a separate paper
\citep{korc2003}. In this paper, we explore the model parameters and
focus on the physical processes of the dynamical evolution of such
infalling dwarfs making this study
beneficial in understanding the properties of merging
parent-satellite galaxy systems.

The numerical realization of this scenario is a complicated problem because of the
large range of masses and sizes of the interacting components,
and due to the drastic change of the mass of the sinking dwarf. We
explain our numerical methods in Sect.~\ref{sec:Numerical_Methods},
where we describe the Galaxy and the dwarf models.
The results of the simulations are presented in
Sec.~\ref{sec:Results}. Some analytical considerations for our results
are given in Sect. 4. Finally we formulate our
conclusions in Sect.~\ref{sec:Conclusions}.

\section{Models and Numerical Methods}
\label{sec:Numerical_Methods}

\subsection{$N$-body simulations}
\label{sec:Nbody_simulations}

In our $N$-body simulations, we neglect gaseous
effects such as pressure and viscosity. We do not include  star formation,
and do not distinguish between visible
and dark matter in the models. Gravitating matter is represented by the softened
particles, and their mutual gravity is calculated by a hierarchical
tree algorithm (see, for example \citet{dubi1996} for a review). The
tolerance parameter which is used to organize the hierarchical boxes,
is 0.7. The softening length is 0.01 of the disc scale length, which
corresponds to 35 pc in our Milky Way model. We use the same value of the softening length 
for all particles irrespective of their mass.

The integration time step is 1/32 of the dynamical time which is
determined as the quarter of the orbital period of a particle at the center of the Milky Way.
This gives a value of $4.38 \times 10^5$ years.
The central density of our dwarf is an order of magnitude larger
than that of the Milky Way model. However, the orbiting time of the stars
in the dwarf is longer compared to the time step.  

An oscillation time of the dwarf  in the potential of the Milky Way is
about 0.5~Gyr at the beginning of an infalling process, and about
0.1~Gyr at later stages of evolution. Since the plane of the orbit of the
dwarf has a low inclination with respect to the disc, 
a characteristic time of the gravitational shocking from the disc
is comparable to the dwarf's orbiting time. The bulge shocking 
time  determined as a time of crossing of the bulge by a moving satellite is much shorter 
(about $3 \times 10^6$ years) compared to the gravitational shoking time from the disk,
making bulge shocking much stronger.

\subsection{The Milky Way Model}
\label{sec:Galaxy_Model}

An accurate model of the Milky Way is important to make a quantitative
estimate of the dwarf evolution. We construct the Milky Way equilibrium model following \citet{kuij1995}.
The model provides wide flexibility in fitting
observational constraints. Briefly, it consists of a nearly
spherical bulge, an exponential disc, and a halo which resembles a
lowered isothermal sphere at larger scales. Each component is expressed
by a distribution function, so that the positions and the velocities
of $N$-body particles can be directly sampled from the distribution
functions. Since the model is close to equilibrium, there is no need to
"relax" it before an actual simulation.

The parameters of the model are chosen to reproduce 
the properties of the real Milky Way galaxy. We choose the
solar radius, the disc exponential scale length ($R_d$), and the disc
vertical scale height ($z_d$) to be 8~kpc, 3.5~kpc, and 245pc respectively. The
circular velocity of the disc at the solar radius is taken to be 220~km/s. The total
surface density within 1.1~kpc of the disc plane, and the contribution
of the disc material to it are 69.8~\msun pc${}^{-2}$, and 45.5~ \msun
pc${}^{-2}$ respectively, which agrees with the observational constraints 
 of $71 \pm 6$ \msun pc${}^{-2}$ and $48 \pm 9$ \msun
pc${}^{-2}$ \citep{kuij1991b}. With a cut-off radius of
28~kpc, the disc total mass is $5\times 10^{10}$~\msun. Toomre's Q value
of the disc is about 1.9 at a distance of 2.5 $R_d$ from the center. 
The bulge mass and the cut-off radius of the bulge
are taken to be $0.75\times 10^{10}$~\msun and 2.38~kpc respectively. 

\begin{figure}
   \includegraphics[width=84mm]{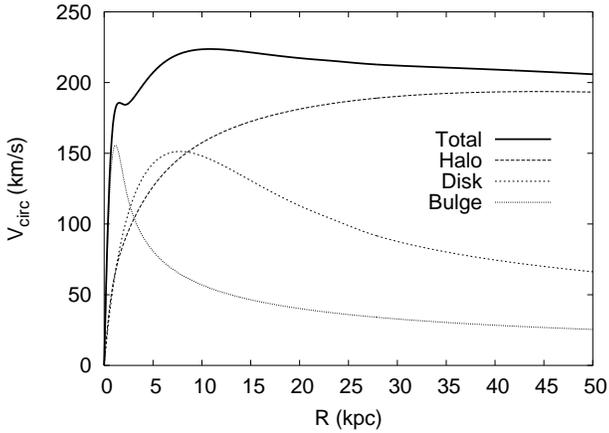}
  \caption{The rotation curve of our Milky Way model. The
    contributions from the bulge, disc and halo to the circular
    velocity are shown by dotted, short-dashed, and long-dashed
    curves, respectively.
     }
  \label{fig:circ_vel}
  \vspace{5mm}
\end{figure}

The halo density distribution can be well approximated 
by a spherical lowered isothermal profile
\citep{binn1987} with a central potential $\Psi(0)/\sigma^2=8$ except
for the central regions which are deformed owing to the presence of the bulge and the disc.
The halo mass inside 50~kpc and 170~kpc radii is  $4.9\times10^{11}$~\msun and
$8.6\times10^{11}$~\msun respectively. This is within the observationally inferred values 
$5.4^{+0.2}_{-3.6} \times 10^{11}$~\msun and $1.9^{+3.6}_{-1.7} \times
10^{12}$~\msun found by Wilkinson and Evans (1999).
The contribution of each component to the circular velocity is shown
in Fig.~\ref{fig:circ_vel}. 

\begin{figure}
  \includegraphics[width=84mm]{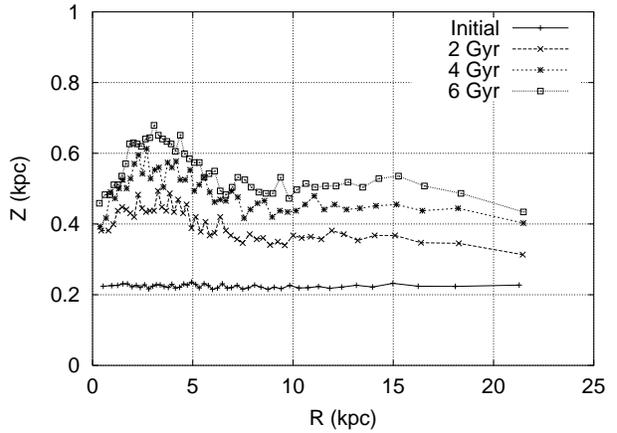}
  \caption{Artificial heating of the Milky Way disc model due to
   two-body relaxation. The number of particles in the bulge and disc
    components is 10,000 and 70,000 respectively, and the halo is
    treated as a fixed potential. The softening length in the tree code is 35~pc.}
  \label{fig:disk_thick0}
\end{figure}

This model is similar to model S in \citet{tsuc2002}. An interested reader
can find in that paper a more detailed description of the model and its stability analysis.
The bulge and the disc in our sumulations are represented by 10,000 and 70,000
equal mass particles. With this rather low number, the disc thickness
grows due to two-body relaxation.
Fig.~\ref{fig:disk_thick0} shows the change in the disc thickness in equilibrium. 
During 6 Gyrs, the disc thickness increases up to 500~pc
near $R=10$~kpc. This is about twice as large as the present day thin disc,
but smaller than the scale height of the thick disc. The bump in the disc thickness
around $R=3$~kpc is caused by a weak bar developing in the disc.
A further discussion of this effect can be found in Tsuchiya (2002). 

In this study, the halo is treated as a fixed potential since we are
mainly interested in the effects of the disc and bulge shocking and
dragging of the dwarf galaxy. We neglect 
dynamical friction from the halo in our models, which  increases 
 the orbital sinking time (see also Section 5.1).
If we treat the halo as an ensemble of $N$ particles,
additional heating causes disc  thickening.
To prevent considerable disc
heating, the number of particles in the halo should be several times
larger than that in the disc. This implies severe limitations on  our computational
capabilities. Nevertheless we run one simulation with a "living"
halo to estimate the contribution of the halo to the dwarf's dynamics. The
halo dynamical friction decreases the dwarf's orbital
decay time and increases its mass loss, but does not change the evolution history
in the mass--apocenter radius plot. We  discuss this issue in section 5.1.

\subsection{Dwarf Models}
\label{sec:Dwarf_galaxy_model}

We assume that a dwarf galaxy captured by the Milky Way was the
hypothetical progenitor of $\omega$ Cen.
Presumably, a  high-density nucleus of the dwarf
survived the disruptive capture event and ended as $\omega$ Cen.
We wish thus to explore dwarf models with 
centrally concentrated density profiles. 
We examine two models: the lowered isothermal or the King model
\citep{binn1987}, and the Hernquist model \citep{hernq1990a}.
Both models are spherically symmetric. 
The kinematics of the compact dwarf spheroidal galaxies does not reveal
a significant rotation of these systems (Binney \& Merrifield 1998).
We assume therefore that both of our models have no rotation.
The King model is defined by the distribution function

\begin{equation}
  f_\mathrm{K}(\varepsilon) = \left\{
    \begin{array}{ll}
      \rho_1(2\pi\sigma^2)^{-3/2}
      \left(e^{\varepsilon/\sigma^2}-1\right) \quad & \varepsilon >0 ;
      \\
      0 & \varepsilon \leq 0,
    \end{array}
    \right.
\end{equation}

where $\varepsilon=-\frac{1}{2}v^2 + \Psi$ is a specific relative energy of the particles
 defined so that the
energy is zero at the surface with zero volume density. The
value of the dwarf potential at the center, $\Psi(0)$, determines
how centrally concentrated the distribution is. We adopt the value $\Psi(0)/\sigma(0)^2 =
12$. 

\begin{figure}
  \includegraphics[width=84mm]{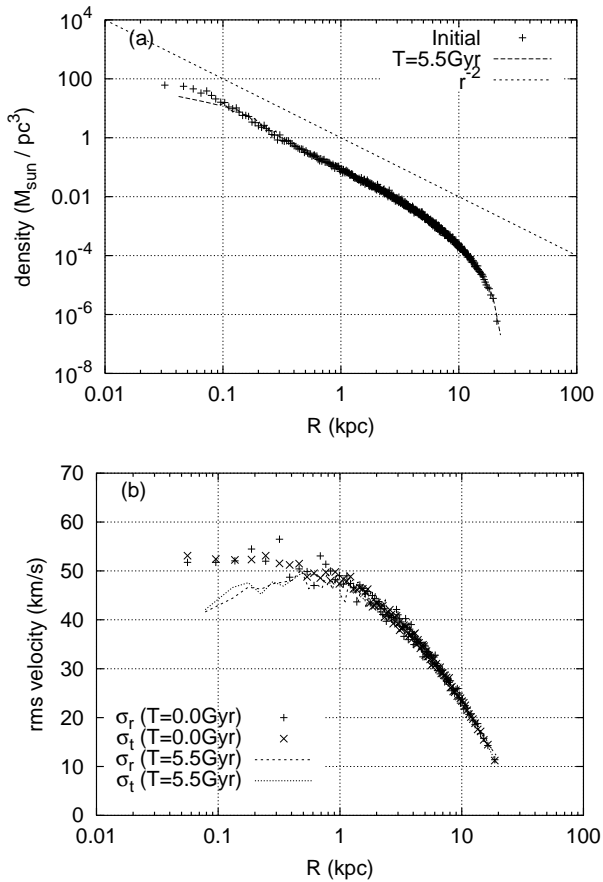}
  \caption{Density distribution (upper panel) and velocity dispersion (lower 
panel) in the King model with
    $\Psi(0)/\sigma^2_0=12$.
    The total mass and the half-mass radius of the model are $8\times 10^9$ \msun and 
    4 kpc respectively (model K5). The distribution is taken from our $N$-body realization with
    $N=50,000$ particles averaged over spherical shells containing 100 particles
    for the density distribution, and 1000 particles for the rms velocity distribution.
    The dashed line in the upper panel shows the density distribution in the model
    after 5.5 Gyr of evolution in isolation. For comparison, 
we also show the $r^{-2}$ density
    distribution (dotted line). 
In the lower panel, the dashed and dotted lines show the radial and 
the tangential
    velocity dispersion profiles respectively, after 5.5 Gyr evolution of the model in isolation.   
    }
  \label{fig:King}
\end{figure}

The model has two characteristic scale lengths: the King
radius $r_K\equiv \sqrt{9\sigma_0^2/4\pi G\rho_0}$, and the tidal
radius $r_t$. For
$\Psi(0)/\sigma_{0}^2=12$, the ratio between the tidal and the King radii,
is $r_t/r_K = 548$. These two scale lengths are not convenient however,
for the comparison with the Hernquist models, so we label the
distributions by their half-mass radii related to the King radius
and the tidal radius as $r_{1/2}=0.164 r_t = 89.9 r_K$. 
Figure \ref{fig:King} shows the density and velocity
profiles in our King model with total mass $8\times10^9$ \msun
and half-mass radius $r_{1/2}=4$~kpc
sampled by 50,000 particles. Since the model has a deep central
potential, a few hundred particles compose the central homogeneous
core of the dwarf. 

To study the numerical stability of the model, we have followed its
evolution in isolation for about 5.5 Gyr. The final distributions
are shown in Fig.~\ref{fig:King}. The central density
decreases by a factor of 2. The change in density profile is noticeable 
within $r<0.1$~kpc of the dwarf center which contains mass of about $10^8$ \msun.
The central velocity dispersion decreases by about 25\%, while there is no considerable change
in the velocity dispersion profile in the outer regions of the dwarf, i. e., at $r>0.5$~kpc.
 
The Hernquist model (Hernquist 1990) has a density profile given by the equation

\begin{equation}
  \rho(r)=\frac{M_\mathrm{tot}r_0}{2\pi r(r+r_0)^3}.
\end{equation}

 In central regions, the density profile
has a cusp of $r^{-1}$, while in the outer regions the density decreases as $r^{-4}$.
The radius $r_0$ separating the two regions is related to the half mass radius as:

\begin{equation}
  r_{1/2}=(1+\sqrt{2})r_0.
\end{equation}

The slope of the density gradient in the central region is thus
shallower than that in the King models. 

\begin{figure}
  \includegraphics[width=84mm]{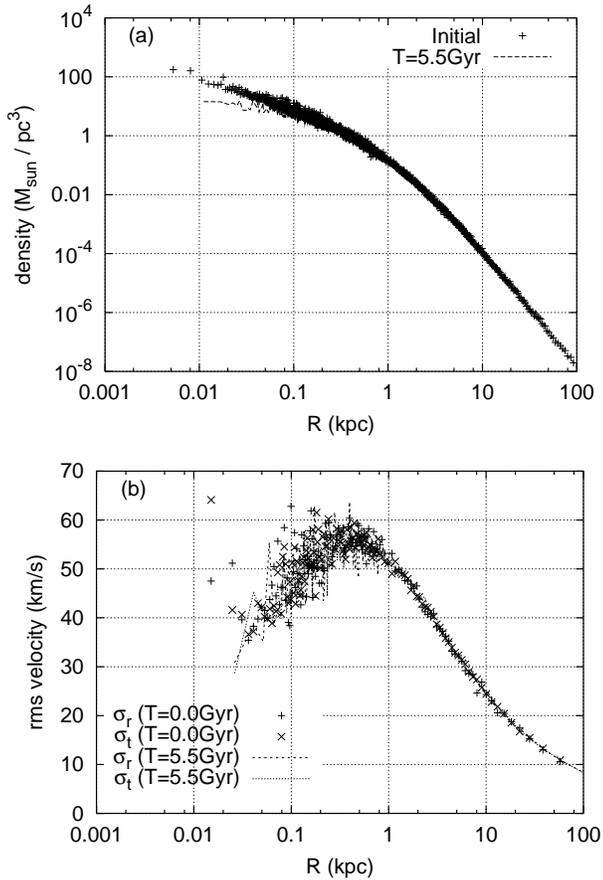}
  \caption{Density (upper panel) and velocity dispersion (lower panel)
distributions for the Hernquist model with total mass of
    $8\times 10^9$ \msun and the half mass radius of 2.83 kpc (model H5) taken
    from multi-mass $N$-body realization with $N=140,000$ particles.
    The density distribution is determined from averages over spherical shells
that contain 100 particles each, while the velocity dispersion distribution
from shells that contain 1000 particles.
    The dashed line in the density plot represents the distribution of the model after 5.5 Gyr of evolution in isolation.
The lines in the velocity dispersion plot show the
radial (dased) and tangential (dotted) velocity dispersions
    after 5.5 Gyr of evolution in isolation.}
  \label{fig:Hernquist}
\end{figure}

A more noticeable difference between
the two models is in their velocity dispersion profiles. While the King
models have constant velocity dispersion in the center, in the Hernquist
models the velocity dispersion decreases near the center as $\sigma^2 \propto r/r_0
\ln(r_0/r)$ \citep{hernq1990a}.
Figure \ref{fig:Hernquist} shows a
realization of the Hernquist model which has a total mass
of $M_\mathrm{tot}=8\times10^9$~\msun and half-mass radius
of $r_{1/2}=2\sqrt{2}$~kpc. This Figure also shows 
the density and the velocity dispersion profiles 
after 5.5 Gyr evolution in isolation. The density in the central region, within the
softening length, decreases in the same way as for the King model. 
However, the velocity dispersion profiles  remain close to the 
initial disrtributions. 

Evolution of the dwarf's central regions is caused by the softening of
the gravity forces. As can be seen from Figure 3,  the velocity dispersion profile
in the King models flattens toward the center, while the volume density 
grows faster than $r^{-2}$. To keep the central regions
in equilibrium, the self-gravity force in the King models should grow towards the center.
The velocity dispersion in equilibruim Hernquist models
decreases toward the center( Figure 4), which
results in a constant self-gravity force in the central regions.
The softening length of 35~pc is comparable to the King radius of 44.5~pc.
Thus, the gravity softening decreases the gravity force in the core regions, and 
causes the core to expand reducing
its density and  velocity dispersion. Central regions of our numerical King equilibrium models 
with softened gravity are more departed from equilibrium than those of Hernquist models,
which results in more noticeable evolution of central regions of the King models.

Decreasing the density at the center of the dwarf can
potentially cause its more rapid disruption. We find however,
that the mass of the King models remains larger than
$10^8$~\msun during the simulations and a structural change
at the dwarf center does not affect mass stripping in the simualtions. 
Our test simulations thus show that isolated dwarf
models are  fairly stable at scales larger than
a few softening lengths. In other words, artificial effects  such as the
softening and the two-body relaxation do not affect the mass stripping
history unless the mass of the remnant falls below  $10^7$ \msun.

\subsection{Multi-Mass $N$-Body Model of the Dwarf}
\label{sec:multi_mass}

With our choice of the softening
length, the numerical resolution near the dwarf's  center is about $10^7$ \msun.
In our King models, we use 50,000 particles to represent
a $8\times 10^9$ \msun dwarf, and a $10^7$ \msun object
would consist of a few tens of particles. 
To improve mass resolution, we increase the number of particles
close to the dwarf center. We find that the Hernquist models  
undergo strong mass loss, and the mass of the remnant becomes less
than $10^7$ \msun during a few Gyr while the King model dwarfs 
stop losing their mass above
$10^8$ \msun. To avoid  effects caused by a small number of particles
at the late stages of evolution, 
a remnant should be composed of more than a thousand particles.
To increase the mass resolution of the Herquist model near the center,
we use a mulyi-mass N-body model of the dwarf. 
We divide the particles in the model into three groups
differing by their energy:

\begin{eqnarray}
  \int_{E \leq E_1} f_\mathrm{H}(E)  d\bmath{x} d\bmath{v} &=& 0.01, \nonumber \\
  \int_{E_1 < E \leq E_2} f_\mathrm{H}(E) d\bmath{x} d\bmath{v} &=& 0.09, \\
  \int_{E_2 < E} f_\mathrm{H} (E) d\bmath{x} d\bmath{v} &=&   0.9,  \nonumber
\end{eqnarray}

where $E=\frac{1}{2}v^2 + \Phi(\bmath{x})$ is a specific energy of the
individual particles. The lowest energy group is sampled by the particles
with mass $m_1 = 2\times 10^{-7} M_\mathrm{tot}$, and  with the number of
the particles in the group equal to 50,000. The middle and the high energy
groups are sampled by the particles with masses $m_2 = 2\times 10^{-6} M_\mathrm{tot}$
and $m_3 = 2\times 10^{-5} M_\mathrm{tot}$ respectively, 
and with the number of particles in each group of 45,000.
Since the light particles are
progressively concentrated near the center, the mass resolution of the dwarf
central regions becomes about a hundred times better.
The central $10^7$ \msun for example is composed now of more than 50,000 particles.

\begin{figure}
  \includegraphics[width=84mm]{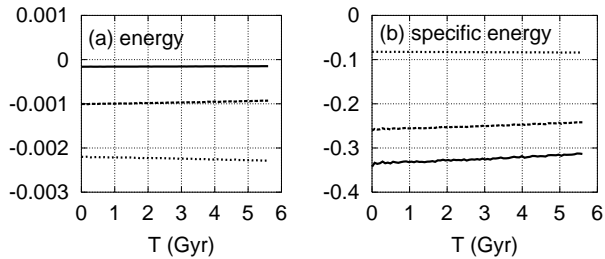}
  \caption{Energy change in the three different mass components of the
    multi-mass Hernquist dwarf. Panel \textbf{(a)} shows the total
    energy of each component, while panel \textbf{(b)} shows the
    specific energy averaged over the particles in each component. The
    solid, dashed, and doted lines stand for the light, medium, and
     heavy mass components respectively.}
  \label{fig:hern_eng0}
\end{figure}

A potential problem of N-body models with different masses of particles
is an artificial heating of light particles. In our model,
light particles are concentrated toward the dwarf's center, while heavy
particles are distributed at the periphery of the dwarf. 
On the contrary, in a dwarf close to equilibrium,
massive particles sink toward its center, while 
light particles are scattered away. In other words,
there is a net energy flow from the massive particles to the light
ones in our model. We have checked whether this process is significant in our simulations.
Fig.~\ref{fig:hern_eng0} shows energy exchange between particles of different masses as
a function of time for the equilibrium multi-mass Hernquist model evolving in isolation.
Solid, dashed and dotted lines in this Figure stand for 
the low, medium and high mass particles. 
The left panel in Fig.~\ref{fig:hern_eng0} shows  the total energy of each 
group of particles as a function of time, while  
the right panel shows the
specific energy, averaged over the particles in each component.
Ninety percent of
the mass of the dwarf in our model is built with massive particles.
Therefore the specific energy of heavy particles does not change
significantly, while the total energy does.  

There is no noticeable
relaxation between the low and intermediate mass particles.
The global energy flow is dominated by the energy transfer from heavy
to the intermediate and low-mass particles with the dominant energy exchange
being between the massive and intermediate mass ones. 
The light particles also gain energy as a result of relaxation.
This energy gain, although, does not cause a significant change in the equilibrium
density ditribution as can be seen in Figure 4.
Furthermore,  heating  of the light particles is even less important
in our simulations since  most of the  heavy particles
are stripped away from the dwarf at early stages of the infalling process.

\subsection{Bound mass}
\label{sec:bound_mass}

The bound mass of a dwarf moving in an
inhomogeneous background field is a rather ambiguous quantity.
One way to determine the dwarf's bound mass is to calculate
the total mass of the dwarf particles that have negative
energy, ignoring the external gravitational field
of the Milky Way: 
\begin{equation}
  \label{eq:sg_mass}
  E_\mathrm{SG}^{(i)}= \frac{1}{2}m_i
  v_i^2+\Phi_\mathrm{dwarf}(\bmath{x}_i) \leq 0.
\end{equation}

Another way is to calculate the mass inside the tidal radius 
where the tidal force from the Milky Way is comparable
to the self-gravity of the dwarf. This would be an appropriate bound mass definition if
the dwarf is subject to a strong tidal field.
To simplify the tidal radius calculations, we assume
that the Milky Way is spherically symmetric
with a homogeneous core of 1-kpc radius, and has a flat rotation curve of 220 km/s
outside the core region. We use the
expression for the tidal radius from Binney and Tremaine (1987)
ignoring the deviation of the Milky Way from spherical symmetry:
\begin{equation}
  r_t=\left(\frac{M_\mathrm{dwarf}(r_t)}{3M_\mathrm{MW}(d)}\right)^{1/3}d,
  \label{eq:tidal_radius}
\end{equation}
Here, $d$ is the distance between the center of the dwarf
(defined at the location of the minimum of the dwarf's potential)
and the center of the Milky Way, $M_\mathrm{dwarf}(r_t)$, and $M_\mathrm{MW}(d)$
are the mass of the dwarf within the tidal radius, and the mass of the Milky Way
within radius $d$.  
Under these assumptions, the dwarf tidal mass is determined
by the expression: 

\begin{equation}
  \label{eq:tidal_mass}
  \frac{M_\mathrm{dwarf}(r_t)}{r_t^3} =\left\{
    \begin{array}{ll}
      \displaystyle
       3.27 \times 10^{10} \Big({V_{rot} \over 220 \mathrm{km} \mathrm{s}^{-1}}\Big)^2 
       \Big({M_{\odot} \over \mathrm{kpc}^3} \Big) & 
      d \leq 1~\mathrm{kpc} \\
      \displaystyle
      {3.27 \times 10^{10} \over (d/1~ \mathrm{kpc})^2} 
      \Big({V_{rot} \over 220 \mathrm{km} \mathrm{s}^{-1}}\Big)^2
      \Big( {M_{\odot} \over \mathrm{kpc}^3} \Big) &
      d > 1~\mathrm{kpc} \\
    \end{array}
  \right.
\end{equation}

As a practical compromise, we also calculate gravitationally bound mass of the dwarf
within its tidal radius. Fig.~\ref{fig:Km08r04_masst} shows the bound mass of an infalling dwarf
as a function of time defined in this way, as well as  the bound mass
calculated with the other two definitions. The bound mass determined 
within tidal radius of the dwarf does not have peculiar features
observed in the mass profile calculated for the mass of the remnant bound by its self-gravity.

\section{Results}
\label{sec:Results}

This section presents the results of the numerical simulations of the infall of a dwarf that has
King and Hernquist density profiles. To match approximately the current orbit of $\omega$ Cen,
the initial position and velocity of the dwarfs have been choosen to be $(X,Y,Z) = (50,0,30)$ kpc,
and $(V_X,V_Y,V_Z) = (0, -20, 0)$ km/s. Here $X,Y,Z$ are the Cartesian coordinates
associated with the Galaxy, $X$ is positive away from the Galactic center, $Y$ is positive
in the direction of the Galactic rotation, and $Z$ is positive toward the North Galactic pole. 
The total masses and the half mass radii of
the examined models are listed in Table 1 for the King models, and in Table 2 for the
Hernquist models. These parameters control the overall evolution of the
dwarf. Massive dwarfs sink faster in the potential well of the Milky Way, while
dwarfs with the large half-mass radii, i.e. a lower volume density 
experience more rapid mass loss.

The King models (K1, K4), (K2, K5) and (K3, K6) listed in Table 1 are related to each other as

\begin{equation}
  \label{eq:selfsimilarK}
  M_\mathrm{dwarf}/r_{1/2}=\mathrm{const}.
\end{equation}  
With the density profile of King models close to $\rho \propto r^{-2}$,
the pairs of models have the same radial density distributions, and implicitly same
central densities. Thus, pair (K1, K4) has the highest central density, while
(K3, K6) has the lowest central density.
The Hernquist models (H1, H4, H7), (H2, H5, H8) and (H3, H6, H9) listed in Table 2, 
are related as
\begin{equation}
  \label{eq:selfsimilarH}
  M_\mathrm{dwarf}/r_{1/2}^2=\mathrm{const}.
\end{equation}
Similarly to the King models, these models have the same density distribution and therefore 
central density within each group.
Group (H1, H4, H7) has the highest central density, while group (H3, H6, H9) has the lowest.

\subsection{King Models}
\label{sec:King}

\begin{table}
  \caption{Parameters of the King models}
  \label{tab:King}
  \begin{tabular}{@{}lcccc}
    \hline
    Model No. &  $M_\mathrm{tot}$ (\msun) & $r_{1/2}$ (kpc) & N & \\
    \hline
    K1 & $4\times 10^9$ & 1.0 & 50000 \\
    K2 & $4\times 10^9$ & 2.0 & 50000 \\
    K3 & $4\times 10^9$ & 4.0 & 50000 \\
    K4 & $8\times 10^9$ & 2.0 & 50000 \\
    K5 & $8\times 10^9$ & 4.0 & 50000 \\
    K6 & $8\times 10^9$ & 8.0 & 50000 \\
    \hline
  \end{tabular}

  \medskip

\end{table}

\begin{figure}
  \includegraphics[width=84mm]{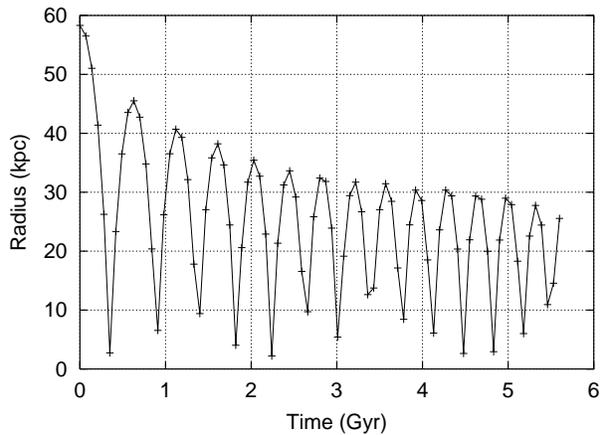}
  \caption{Time evolution of the orbital radius of the dwarf model K5.}
  \label{fig:Km08r04_radt}
\end{figure}
All the King models listed in Table 1 are composed of 50,000 identical particles
and have central potential depth of $\Psi(0)/\sigma^2_0=12$. 

Figure 6 shows the time evolution of the dwarf orbit for model K5  
taken as an example. This plot shows the distance of the dwarf from the center of the Milky Way
as a function of time
with the position of the dwarf determined as the point where the dwarf potential reaches its
minimum.

The dwarf has a highly eccentric orbit which decays
due to the interaction with the disk and the bulge of the Milky Way.
The distance minima in Figure 6 do not correspond to the exact pericentric passages of the dwarf
because of the rather low time resolution in the plot, that is about 700 Myr.
We estimate the typical dwarf's pericentric distances of about 1~kpc. 

\begin{figure}
  \includegraphics[width=84mm]{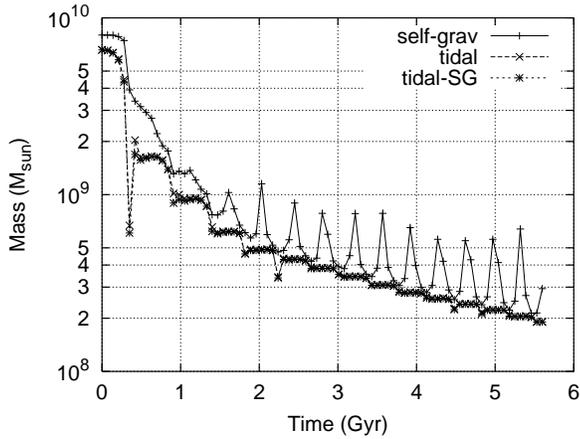}
  \caption{The mass of the dwarf model K5 as a function of time. The
    solid and dashed lines show the mass of the dwarf calculated as gravitationally bound and tidal
    mass respectively. The dotted line is the bound mass 
    determined as  gravitationally bound mass of the dwarf within its tidal radius.}
  \label{fig:Km08r04_masst}
\end{figure}

Fig.~\ref{fig:Km08r04_masst} shows the mass loss history for the dwarf model K5.
We use three different definitions of the dwarf's bound mass.
The dashed line shows the mass of the dwarf within its tidal radius.
The dotted line shows the behavior of the gravitationally bound mass of the dwarf
within its tidal radius.
For comparison, we plot also the
mass of the gravitationally bound particles of the whole dwarf (solid line).
The mass of the remnant determined by the gravitationally bound particles is
larger than the mass of the remnant within its tidal radius.
Until  t $\sim 1.4$ Gyr, the mass of gravitationally bound particles of the dwarf
decreases monotonically, changing later on into an oscillatory regime.
By defining the remnant mass as the mass of the dwarf's gravitationally bound particles,
we neglect the gravity of the Galaxy. With this definition, the particles
with sufficiently small velocities will be gravitationaly bound even if they
are located at large distances from the dwarf center. Close to apocenter,
the dwarf and the stripped particles slow down, which results in an increase
of the dwarf bound mass. We find this definition inappropriate for our problem.

The mass of the remnant determined by its tidal radius changes mainly
during its pericentric passages and remains nearly constant for the rest of the orbiting time.
A simple approximation for the tidal radius estimate is
invalid however close to pericenter, where the dwarf's distance
from the Galactic center is comparable to the size of the dwarf itself.
The bound mass determined as a mass of gravitationally bound particles
within the tidal radius ofthe dwarf is thus the most plausible
definition of the gravitationally bound mass for our problem.
 
The orbital sinking of the dwarf is caused by dynamical friction, or a
back reaction from the disc and bulge density perturbations
determined by the dwarf's gravitaty. 
In an ideal case, the drag force is proportional
to the square of the dwarf mass (Chandrasekhar 1943). As the dwarf loses its mass, the
orbital decay rate decreases. The history of the orbital decay
depends thus crucially on the mass loss rate of the dwarf, which in turn
depends on dwarf's internal density distribution.
We study these complex connections in an evolutionary history of the dwarf 
by a systematic parameter survey of our models.

\begin{figure}
  \includegraphics[width=84mm]{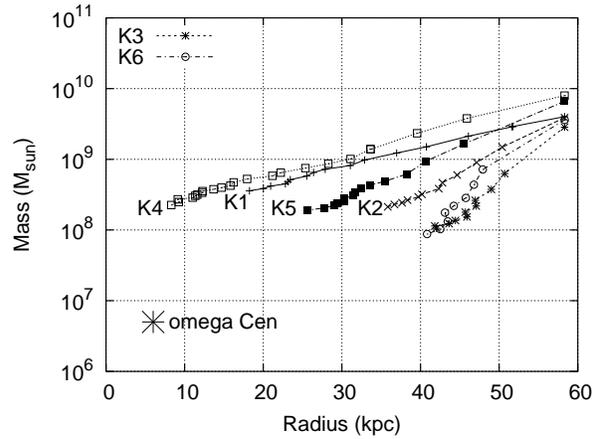}
  \caption{Evolution of the mass as a function of apocentric
    distance for the King models. The labels in Figure match
    the models listed in Table~\ref{tab:King}.}
  \label{fig:Kmassrad}
\end{figure}

An illuminating diagram is Figure 8 which shows the bound mass as a function of
apocentric distance of the dwarf. 
The evolution history of King dwarf models can be classified into the three
groups depending on their initial central density. Models in each pair, e.g, (K1,
K4), (K2, K5), and (K3, K6) reach nearly the same bound mass and
apocentric distance in their evolutionary
courses. These model pairs have the same density distributions,
and once the outer parts of the dwarf are stripped, the 
models in each pairs have nearly identical remnants.
 The
bound mass of the King model dwarfs decreases exponentially with apocentric radius. 
An extrapolation of the evolution of model K2 could lead in principle
to a remnant with the mass and the orbital parameters close to those of
$\omega$ Cen. However, the evolution time to reach
$\omega$ Cen's parameters is larger than the age of the Galaxy,
and $\omega$ Cen's origin can not be explained by this model.

\begin{figure}
  \includegraphics[width=84mm]{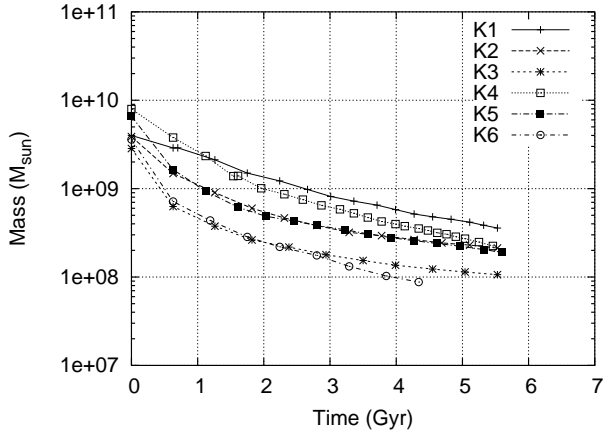}
  \caption{Evolution of the mass of the King dwarfs as a function of time.}
  \label{fig:Kmasst}
\end{figure}

Figure 8 also illustrates a common evolution of the models 
that have the same central densities.
Models K2 and K5, for example, have close masses
by the second pericentric passages. Models K3 and K6 have close
bound masses even at the beginning of the simulations since this pair
of models has the lowest central density, and the tidal radius for these models is already
small at $R \sim 60$~kpc. The additional initial mass in model K6 compared to the
model K3 is distributed outside the tidal radius (as defined in Section 2.5), 
and is stripped during the very first passage by the Galactic center.

Figure 9 illustrates the mass evolution of the King dwarf models. By 6 Gyr,
the most the dwarf models reach mass of about few$\times 10^8$ M$_{\odot}$,
with the mass loss comparable with that found in Zhao' model (Zhao 2002).
Although the models K3 and K6 have higher mass loss rates and reach 
10$^8$ M$_{\odot}$ during 4 Gyr of evolution.

\begin{figure}
  \includegraphics[width=84mm]{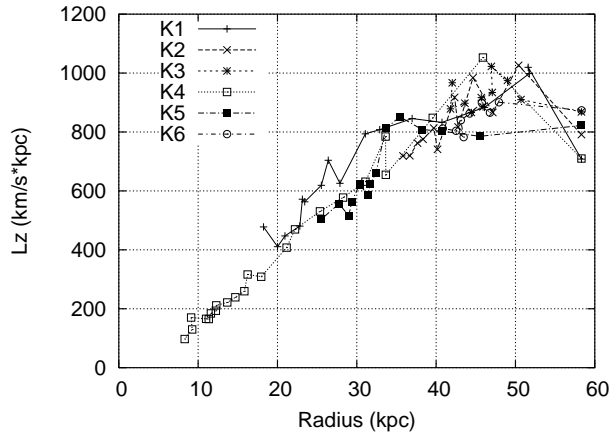}
  \caption{Evolution of the angular momentum 
    as a function of the decaying apocentric radius of the King dwarfs.}
  \label{fig:Klzrad}
\end{figure}

A similar evolutionary pattern among different King models can be
seen also in dependence of the specific angular momentum $L_z$. 
Fig.~\ref{fig:Klzrad} shows $Lz$ with
the apocentric radius. The positive $L_z$ is for retrograde orbits
to the disc rotation. At the beginning, we set the same orbital
parameters for all models so that all models have the same
initial $L_z$. 

\subsection{Hernquist Models}
\label{sec:Hernquist}

\begin{table}
  \caption{Parameters of the Hernquist models}
  \label{tab:Hernquist}
  \begin{tabular}{@{}lccc}
    \hline
    Model No. &  $M_\mathrm{tot}$ (\msun) & $r_{1/2}$ (kpc) & N \\
    \hline
    H1 & $4\times 10^9$ & 1.0 & 140000 \\
    H2 & $4\times 10^9$ & 2.0 & 140000 \\
    H3 & $4\times 10^9$ & 4.0 & 140000 \\
    H4 & $8\times 10^9$ & 1.414 & 140000 \\
    H5 & $8\times 10^9$ & 2.828 & 140000 \\
    H6 & $8\times 10^9$ & 5.657 & 140000 \\
    H7 & $16\times 10^9$ & 2.0 & 140000 \\
    H8 & $16\times 10^9$ & 4.0 & 140000 \\
    H9 & $16\times 10^9$ & 8.0 & 140000 \\
    \hline
  \end{tabular}

  \medskip
\end{table}

To examine the dynamics of the Hernquist model dwarfs, we build nine models
listed in Table 2.
As it was mentioned before, there are three groups of models namely 
(H1, H4, H7), (H2, H5, H8), and (H3, H6, H9)  which differ
in their half-mass radii and  mass but have the same radial density distributions
(i. e., central densities).

\begin{figure}
  \includegraphics[width=84mm]{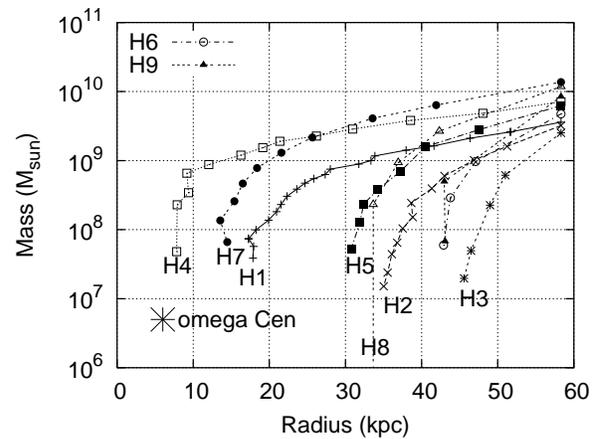}
  \caption{Evolution of the mass of Hernquist dwarf models as a function
    of dwarf's apocentric radius. }
  \label{fig:Hmassrad}
\end{figure}
Figure 11 shows mass evolutionary tracks for the nine Hernquist models.
In this Figure, the bound mass of the remnant is plotted as a function of
its apocentric distance. Similarly to the King models, a common behavior of the 
Herquist models which have the central density
is seen in Figure 11. Once the
tidal interaction strips the outer layers of a dwarf belonging
to the same group, the remnants become nearly identical
and have close orbital behavior.

A difference in the behavior of the Hernquist models
when compared to the King models is the bending of the evolutionary tracks
at lower masses. The decay
of the apcentric radius stops in both models, but the Hernquist model dwarfs continue to loose 
their mass after the orbit evolution is terminated. 
All of the models but H1 are completely disrupted
in our simulations during a finite time. 

The central mass distribution of a dwarf can not always be represented
by the Hernquist profile. The center of a dwarf may have a much denser nucleus
or even a black hole. In this case, the dense central parts
of the dwarf may survive the disruptive gravitational shocks
and settle on a low-energy orbit of apocentric radius smaller than the Solar circle 
for instance, or even settle within the bulge of the Galaxy.
Our simulations follow the mass evolution of the dwarf until the mass
of the remnant is about few $\times 10^7$ \msun. As it is seen from  
Fig.~\ref{fig:Hmassrad}, the dwarf orbits stop their decay
when the mass of the remnant is below $10^8$ \msun. 
Therefore, regardless the mass
evolution of the dwarf at the very final stages, 
a hypothetical central object will be launched at the dwarf orbit
when mass of the remnant is about $5\times 10^7$ \msun. 
The parameters of the final orbit of the
central nucleus of the dwarf can be estimated rather firmly. 
In particular, the central nucleus of the Hernquist dwarf which has the high volume density 
will have an orbit with apocentric distance of $10\sim20$~kpc. 
Remarkably, model H4 sinks to the Galactic center with apocentric radius of 6~kpc
which is close to an apocentric distance of $\omega$ Cen.
It has to be noticed, however, that a progenitor galaxy has to contain
in its center a comparact self-gravitating nucleus which 
can survive disruptive encounters with the bulge and the disk of the Milky Way. 

The disk thickening due to an artificial heating, and due to an interaction
with the satellite can affect the orbit decay. This effect depends though
 on the satellite orbit.
In our simulations which model the origin of Omega Centauri,
the orbit decay occurs during first 2 Gyr, or during five-six dwarf
revolutions around the Milky Way center (Tsuchiya et al. 2003).
We find that during this time, the satellite orbit is not co-planar
with the Milky Way disk, and the dwarf spends a short time
inside the finite thickness disk of the Milky Way. 
In other words, the disk remains 'thin' for the dwarf,     
and the gravitational drag force acting at every point of the dwarf
orbit does not change much compared to the drag force for the  
disk with the fixed thickness.

The final apocentric distance of the nucleus is not a monotonic 
function of the initial mass of the dwarf.
The model with the smallest initial mass 
H1 ends with an apocentric radius of 18~kpc, while the models with larger initial
masses end their course closer (H4) as well as further (H8) 
to the Galactic center. 

\begin{figure}
  \includegraphics[width=84mm]{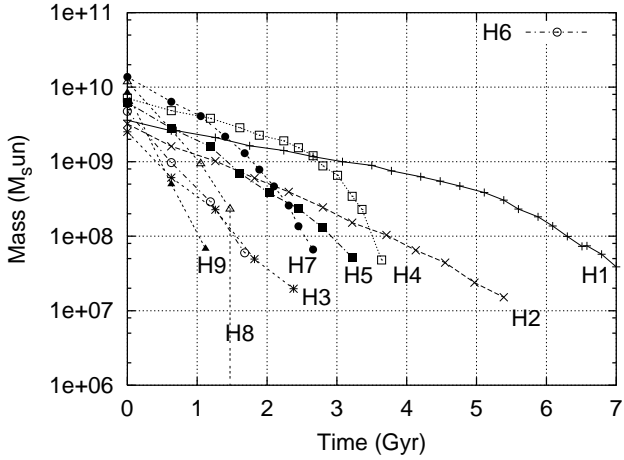}
  \caption{Evolution of the bound mass of the Hernquist dwarfs as a
    function of time.}
  \label{fig:Hmasst}
\end{figure}

Fig.~\ref{fig:Hmasst} shows the evolution of the bound mass as a function of time
which is plotted at the apocenters. All models except 
model H1 have truncated evolutionary curves with the dwarf completely
disrupted in a finite computational time.

\begin{figure}
  \includegraphics[width=84mm]{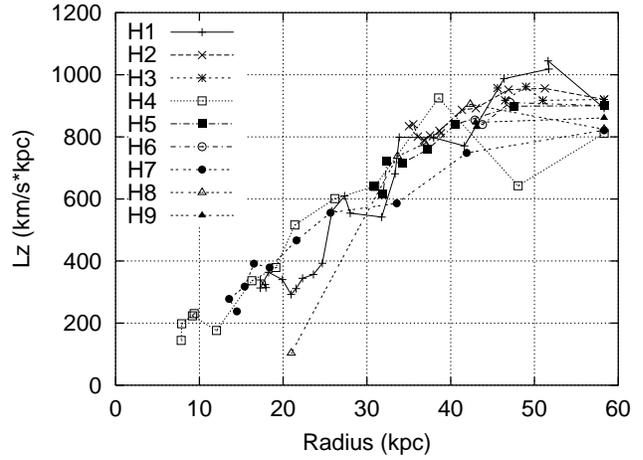}
  \caption{Evolution of the angular momentum 
    as a function of the decaying apocentric radius of the Hernquist dwarfs.}
  \label{fig:Hlzrad}
\end{figure}

Figure 13 shows the specific angular momentum as a function of
apocentric distance of the sinking dwarf.
The evolution of the specific angular momentum of the Hernquist models is
not very different from that of the King models.

\section{Hernquist Models vs King Models}
\label{sec:difference}

We see from our simulations a difference in the behavior of the King and
the Hernquist models. The mass loss rate for the King model dwarfs 
slows down as the bound mass decreases, while for the Hernquist model
dwarfs the mass loss rate increase.
The Hernquist model dwarfs are more fragile than the King
dwarfs regardless of their mass and size because they have
shallower central concentration. This, however, only partly
explains the fragility of the Hernquist models. 
We present in this
section simple arguments explaining the difference between
the King and the Hernquist models.

There are two factors that affect central density distribution
of the dwarfs.
The gravitational shocks inject energy into the dwarf's central regions.
Another factor is an indirect influence of gravitational shocks 
on the density distribution in the dwarf's central regions.
The central density distribution changes as a response to
the loss of the outer layers of the dwarf when the dwarf
adjusts itself to a new equilibrium. Although these two effects
act simultaneously, we estimate them separately 
in our simple analytical approach. 

At the later stages of its evolution, the dwarf orbit has a low 
inclination toward the disc and the
bulge shocking is larger than that from the disc. 
An approximate expression for the energy injection due to a single
bulge shocking (Spitzer 1958) has the form:
\begin{equation}
  \Delta E = \frac{M_\mathrm{dwarf} r_m^2}{3} \left(
    \frac{2GM_\mathrm{bulge}}{p^2 V} \right)^2 L(\beta)
  \label{eq:shock}
\end{equation}
Here $p$, $V$, $r_m$, and $\beta$ are the pericentric distance ($\sim$
1~kpc), the relative velocity ($\sim$ 300~km/s), the mean radius of the
dwarf and the ratio of
the duration of the shock to the stellar orbital period
\begin{equation}
  \beta = \frac{p}{V} \frac{v_\mathrm{rms}(r)}{r}.
\end{equation}
The correction factor to the impulsive approximation, $L(\beta)$, is
determined by the relative interaction time $\beta$ such that:
\begin{equation}
  L(\beta)=1, \quad \mathrm{for} \quad \beta\ll 1;  \quad
  L(\beta)\ll1, \quad \mathrm{for} \quad \beta\gg 1.
\end{equation}
Equation (10) can be further approximated by 
replacing the dependence on $M_\mathrm{dwarf}$ and $r_m^2$ in eq.~(\ref{eq:shock}) by
that of the mass of the dwarf within radius $r$.

The central density distributions for the King and the Hernquist models
can be well approximated by a power-law distribution
\begin{equation}
  \rho(r) = \rho_0 \left(\frac{r}{r_0}\right)^n,
  \quad n=-2 \mbox{ or } -1,
\end{equation}
This profile is a
simplification of our King model ($n=-2$) and the Hernquist model
($n=-1$). Calculating gravitational potential energy for the density
distributions (13), we can find
a binding energy of the central sphere of radius $r$ for the
King and the Hernquist models respectively:
\begin{equation}
  E_\mathrm{bound_K}= \frac{GM(r)^2}{2r},
\end{equation}
and
\begin{equation}
  E_\mathrm{bound_H}= \frac{GM(r)^2}{3r},
\end{equation}
With help of the above expressions,  
 one can write down the fractional energy injection
for the King model as:
\begin{equation}
  \frac{\Delta E}{E_\mathrm{bound_K}} =
  \frac{16}{3}\frac{M_\mathrm{bulge}}{M_\mathtt{tot}}
 \left( \frac{r_\mathrm{h}}{r} \right) \frac{r^3}{p^3}
  \frac{GM_\mathrm{bulge}}{pV^2} L(\beta) {\mbox ,}
\end{equation} 
Here, the mass of the King central sphere of radius $r$
has been expressed using total mass $M_{tot}$, and the half mass
radius $r_h$: $M(r) = M_{tot}(r/2r_h)$. 
This gives, for example, for the King model K4:
\begin{equation}
 \frac{\Delta E}{E_\mathrm{bound_K}} =
3.7 \left( \frac{r}{1\mathrm{kpc}}\right)^2 L(\beta)
\end{equation} {\mbox .}
Similarly, we find for the Hernquist model H4
\begin{equation}
\frac{\Delta E}{E_\mathrm{bound_H}} =
5.5 \left( \frac{r}{1\mathrm{kpc}}\right) L(\beta)
\end{equation}
For the accepted parameters, the energy injection 
is smaller than the binding energy if $r < 0.5$ kpc for the King
model, and if $r < 0.2$ kpc for the Hernquist model.

More important effect is a subsequent adjustment of the remnant to a new
equilibrium. 
Let's assume that a Hernquist model is instantaneously stripped to a 
cutoff radius $r_{cut} < r_0$.
In this simplified model we consider mass stripping
as the decrease of the cutoff radius $r_{cut}$.
Using the Jeans equation, we can write the expression
determining the velocity dispersion of the dwarf at a central region $r$ as a function
of cutoff radius $r_{cut}$:

\begin{equation}
  \sigma^2(r{\mbox ,} r_{cut}) = \left\{
    \begin{array}{ll}
      2\pi G \rho_0 r_0^2\left(1-r^2/r_{cut}^2\right),\quad & n=-2,\\

      2\pi G \rho_0 r_0 r \ln r_{cut}/r, & n=-1.
    \end{array}
    \right.
\end{equation}

For the King profile, the
central velocity dispersion does not depend much on the cutoff radius $r_{cut}$, 
while for the Hernquist profile,
the equilibrium velocity dispersion should be lower for
smaller $r_{cut}$.
The King model dwarfs keep thus their
central density distribution almost intact. The inner regions are close to
equilibrium even though the outer layers are removed.
On the other hand, the Hernquist model dwarf expands after the shock
because it has an excess of internal kinetic energy
necessary for equilibrium. This would decrease the density
of the dwarf, making it more fragile to future gravitational shocks
which leads to a complete disruption of the dwarf in  a relatively short
time.   

\section{Discussion}

\subsection{Living halo}
\label{sec:live_halo}

In all of our simulations presented in this paper we assumed that the halo has
a fixed potential. This assumption does not allow us to take into account the drag force due to
halo dynamical friction that can change the dynamics
of the dwarf. For comparison, we ran one simulation with a living
halo for dwarf model H4. The halo gravity was calculated with
the SCF algorithm \citep{hernq1992}. The halo is sampled
by 100,000 particles, and the self-consistent dynamics of the whole
system is thus treated by a hybrid algorithm, SCF-TREE
\citep{vine1998,tsuc2002}.
\begin{figure}
  \includegraphics[width=84mm]{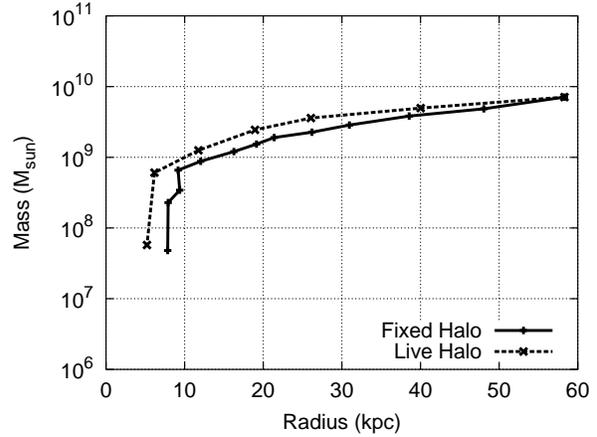}
  \caption{Comparison of the evolutionary histories between models
  with and without the halo responsiveness. In the apocenter -- bound-mass 
 diagram both models take nearly the same track.}
  \label{fig:livehalo}
\end{figure}
As expected, we find  that the orbit decays faster, and
the mass loss rate is higher when compared to that of the models with the fixed halo. 
For the fixed halo model,  the apocenter of the H4-dwarf becomes smaller than 10~kpc
after 3 Gyr, while in the living-halo simulation it happens after 1.8~Gyr.
However, the evolutionary tracks for the fixed and living halo models 
are remarkably similar in the bound mass -- apocentric distance diagram
(Fig.~\ref{fig:livehalo}).
The shock strength depends on the pericentric passage velocity, which in turn determines
the apocentric distance of the dwarf orbit. As a result, the bound mass - apocentric distance
evolutionary tracks are nearly independent of halo model.
Since our basic conclusions are based on the bound mass --  apocentric
distance diagrams, we correctly predict the final orbit of the remnant.

\subsection{$\omega$ Centauri}
\label{sec:omega_cen}

An important application of our simulations is the understanding of the origin of
$\omega$ Cen.  As discussed in Section 4, an
$\omega$ Cen -- like object cannot be created from a dwarf with a highly concentrated King
density profile. On the other hand, the dwarfs with Hernquist density profiles exhibit
accelerated mass loss and are completely disrupted in a few Gyr.
A solution  to this dilemma might be the existence of a dense compact
nucleus at the center of the Hernquist dwarf. 
Such a dense compact nucleus can sink together with the disrupting,
decaying dwarf and may survive the 
disruptive encounters with the disc and bulge of the Milky Way.
A nucleus with a mass of about  $10^7$\msun does not affect the density
distribution of the dwarf,  and most of the evolutionary history
will be similar to that of a non-nucleated dwarf galaxy.
This scenario has been examined in \citet{korc2003}. We have placed a massive
particle of $10^7$\msun at the center of the Hernquist model H4, and followed
its self-consistent evolution. The nucleus
settles into an orbit with apocenter of about 6~kpc, and has a bound mass
approaching $10^7$\msun. A nucleus as a single massive particle is obviously a simplification
of the model which does not allow us to model realisticallly the dynamics of the stripped
nucleus of the dwarf. Although, if the N-body nucleus is compact enough, e.g. 
can be represented by a King model with a core radius of $\sim 10$ pc, such 
a nucleus will survive for a few Gyr on its present orbit.
Nucleated dwarf galaxies are common among dwarf
galaxies, and such a scenario seems to be a plausible explanation
of the origin of $\omega$ Cen.

We arbitrarily modeled a compact nucleus by a softened particle with a 35 pc 
half mass radius, equal to the gravity softening length.
Zhao (2003) noticed recently that such a nucleus is more fluffy than the internal 
central density
concentration of the Hernquist model, and might be destroyed by the
tidal forces from the surrounding dwarf.
A more compact nucleus would obviously remedy the problem.
However, with our 35 pc resolution, modeling of a more compact
nucleus would not make any difference. 
Additionally, a compact nucleus will not affect an orbital and
mass history of the satellite. 
We note also that 
the half-light radius of $\omega$ Cen is 4.8$^{\prime}$ (de Marchi, 1999) 
or about 7 pc. 

The subsequent evolution of the proto - Omega Cen object is determined by a number of
processes: dynamical friction, evaporation driven by two-body relaxation,
gravitational shocks, and mass loss driven by stellar
evolution (Fall \& Zhang 2001).
Our numerical resolution and the assumed model do not allow us to discuss
other effects that might affect the future evolution of the remaining object.
Gnedin et al. (2002) estimate the mass loss from $\omega$ Cen due to
the stellar evolution to be $10^{-2} M_{cl}$ Gyr$^{-1}$. Gravitational shocks
which dominate the mass evolution of the massive globular clusters 
also have a time scale of about a few Gyr (Fall \& Zhang 2001). 
The dynamical friction is negligible for $\omega$ Cen.
We can speculate therefore that an object which has mass close to
that of $\omega$ Cen and which has a compact enough density distribution
can survive on its orbit during a few Gyr.

Recently, Bekki and Freeman (2003) have discussed a similar model of the origin
of $\omega$ Cen. They modeled the merger of a nucleated dwarf galaxy
with an ancient bulgeless Galaxy whose thin disk has twenty percent
of today's thin-disk mass, and included effects of star formation in the dwarf dynamics.
They find that the Galactic tidal force induces radial inflow in the center 
of the dwarf that triggers nuclear starbursts. Bekki and Freeman (2003)
conclude therefore that the effecient dwarf nuclear chemical enrichment can be 
associated with the origin of the observed metal-rich stars in $\omega$ Cen. 
\section{Conclusions}
\label{sec:Conclusions}

We have studied the dynamical evolution of a falling dwarf galaxy,
which in turn is influenced by the strong gravitational shocks from the bulge and disc of the host galaxy.
We focused our study on the mass-loss history, and on the orbital migration of the dwarf. 
The dwarf galaxy sinks toward the center of the host galaxy 
due to dynamical friction, which depends on the
dwarf's mass. Simultaneously, the dwarf is losing mass
due to strong bulge and disc gravitational shocks. As a result, the mass-loss history 
and orbital decay are closely connected and depend in turn on the
internal density distribution of the dwarf.
We have examined two different dwarf models, namely a highly-concentrated King model
and  the Hernquist model. The Hernquist model has a $r^{-1}$ density cusp while the
King model has a density profile nearly proportional to $r^{-2}$. This
difference leads to qualitatively different evolutionary courses
for the two models.
Regardless of the mass and the size, all the King models stop
orbital sinking in a few Gyr, once the remnant mass
has reached $\sim 10^8$ \msun. The central regions of the King model dwarf
remain gravitationally bound, and the remnant continues to orbit for a long
time without considerable mass loss. Therefore, if the initial density distribution
of the dwarf galaxy is close to isothermal, its remnant can not
have a mass much smaller than $10^8$ \msun.
On the other hand, the Hernquist model dwarfs keep losing mass
even after dynamical friction becomes ineffective. As a
consequence, the remnant can be stripped down to a small object. If there is a
compact gravitationally  bound nucleus in the center of the dwarf with a
mass of $\la 10^7$ \msun and will the size of a globular cluster, it will
survive for few Gyr on the present orbit of $\omega$ Centauri (e.g., Fall \& Zhang 2001).
Our results support the capture scenario for the origin of
the globular cluster $\omega$ Centauri.

\section*{Acknowledgments}

The authors thank the referee, M. Wilkinson, for his comments and
questions which helped to improve the paper.
The authors thank W. van Altena and T. Girard for their comments on the manuscript. 
TT and VK are grateful for stimulating discussions with Rainer Spurzem, Andreas
Just and Michael Fellhauer. TT was supported by Alexander von Humboldt
foundation. VK was partly supported by DFG grant 436 RUS 17/112/02.


\begin{thebibliography}{19}
\expandafter\ifx\csname natexlab\endcsname\relax\def\natexlab#1{#1}\fi
\expandafter\ifx\csname url\endcsname\relax
  \def\url#1{\texttt{#1}}\fi
\expandafter\ifx\csname urlprefix\endcsname\relax\def\urlprefix{URL }\fi

\bibitem[{{Bekki} and {Freeman}(2003)}]{bf2003}
{Bekki} K., {Freeman} K.C., 2003, MNRAS, 346, L11

\bibitem[{{Binney} and {Tremaine}(1987)}]{binn1987}
{Binney} J., {Tremaine} S., 1987, "Galactic dynamics". Princeton University
  Press, Princeton NJ

\bibitem[{{Binney} and {Merrifield}(1987)}]{binn1998}
{Binney} J., {Merrifield} M., 1987, "Galactic astronomy". Princeton University
  Press, Princeton NJ 

\bibitem[{Chandrasekhar(1943)}]{chand1943}
Chandrasekhar, S. 1943, ApJ, 97, 255

\bibitem[{De Marchi(1999)}]{demar1999}
De Marchi, G., 1999, AJ, 117, 303

\bibitem[{Dinescu et~al.(1999)Dinescu, Girard, and {van Altena}}]{dine1999}
Dinescu D.~I., Girard T.~M., {van Altena} W.~F., 1999,  AJ, 117, 1792

\bibitem[{{Dubinski}(1996)}]{dubi1996}
{Dubinski} J., 1996, New Astronomy, 1, 133

\bibitem[{Fall(2001)}]{fall2001}
Fall, Michael S., Zhang, Quing, 2001, ApJ, 561, 751

\bibitem[{Freeman(2001)}]{free2001}
Freeman K.~C., 2001,  In: Deiters S., Fuchs, B., Just, A., Spurzem, R.,
  Wielen, R. (Eds.), ``Dynamics of Star Clusters and the Milky Way'', p.~43

\bibitem[{Freeman(1993)}]{free1993}
Freeman, K. 1993, in IAU Symp. 153, Galactic Bulges, ed. H. Dejonghe \& H. J.
    Habing (Dordrecht: Kluwer), 263

\bibitem[{Freeman(2002)}]{freeman2002}
Freeman, K., Bland-Hawthorn, J., 2002, ARA, 40, 487

\bibitem[{Gnedin (2002)}]{gned2002}
Gnedin, O. Y., Zhao, H. S., Pringle, J. E., Fall, S. M., Livio, M.,\& Meylan, G.
    2002, ApJ, 568, L23


\bibitem[{Hernquist(1990)}]{hernq1990a}
Hernquist L., 1990, ApJ, 356, 359

\bibitem[{Hernquist and Ostriker(1992)}]{hernq1992}
Hernquist L., Ostriker J.~P., 1992, ApJ, 386, 375

\bibitem[{Ibata et al.(2001)}]{ibata2001}
Ibata, R., Irwin, M., Lewis, G.,\& Stolte, A. 2001b, ApJ, 547, L133

\bibitem[{Kuijken and Dubinski(1995)}]{kuij1995}
Kuijken K., Dubinski J., 1995, MNRAS, 277, 1341

\bibitem[{{Kuijken} and {Gilmore}(1991)}]{kuij1991b}
{Kuijken} K., {Gilmore} G., 1991, ApJL, 367, L9

\bibitem[{Searle and Zinn(1978)}]{searle1978}
Searle, L., \& Zinn, R. 1978, ApJ, 225, 357 

\bibitem[{Smith (2000)}]{smith2000}
Smith, V. V., Suntzeff, N. B., Cuhna, K., Gallino, R., Busso, M., Lambert, D. L.,
   \& Straniero, O. 2000, AJ, 119, 1239 

\bibitem[{{Spitzer, Jr.}(1958)}]{spit1958}
{Spitzer, Jr.} L., 1958, ApJ, 127, 17

\bibitem[{Tsuchiya(2002)}]{tsuc2002}
Tsuchiya T., 2002, NewA 7, 293

\bibitem[{Tsuchiya et~al.(2003)Tsuchiya, Dinescu and Korchagin(2003)}]{korc2003}
Tsuchiya T., Dinescu D., Korchagin, V., 2003, ApJL, 589, L29


\bibitem[{Vanture(2002)}]{vanture2002}
Vanture, A. D., Wallerstein, G.,\& Suntzeff, N. B. 2002, ApJ, 569, 98

\bibitem[{Vine and Sigurdsson(1998)}]{vine1998}
Vine S., Sigurdsson S., 1998, MNRAS, 295, 475

\bibitem[{{Wilkinson} and {Evans}(1999)}]{wilk1999}
{Wilkinson} M.~I., {Evans} N.~W., 1999, MNRAS, 310, 645

\bibitem[{Zhao(2002)}]{zhao2002}
Zhao H.S. 2002, in ASP Conf. Ser. 265,  Centauri: A Unique Window into
    Astrophysics, ed. F. van Leeuwen, J. D. Hughes, \& G. Piotto (San Francisco:
    ASP), 391

\bibitem[{Zhao(2003)}]{zhao2003}
Zhao H.S. 2003, MNRAS, submitted

\end{thebibliography}

\bsp

\label{lastpage}

\end{document}